# New directions in mechanism design

Jiangtao Meng *
globangrilion@yahoo.com

July 6, 2018

**Abstract**

Mechanism design uses the tools of economics and game theory to design rules of interaction for economic transactions that will,in principle, yield some desired outcome. In the last few years this field has received much interest of researchers in computer science, especially with the Internet developing as a platform for communications and connections among enormous numbers of computers and humans. Arguably the most positive result in mechanism design is truthful and there are only one general truthful mechanisms so far : the generalized Vickrey-Clarke-Groves (VCG) mechanism. But VCG mechanism has one shortage:"the implementation of truthfulness is on the cost of decreasing the revenue of the mechanism." (e.g., Ning Chen and Hong Zhu. [1999]). We introduce three new characters of mechanism:partly truthful, critical , consistent, and introduce a new mechanism: X mechanism that satisfy the above three characters. Like VCG mechanism, X mechanism also generalizes from Vickery Auction and is consistent with Vickery auction in many ways, but the extended methods used in X mechanism is different from that in VCG mechanism . This paper will demonstrate that X mechanism better than VCG mechanism in optimizing utility of mechanism , which is the original intention of mechanism design. So partly truthful,critical and consistent are at least as important as truthful in mechanism design, and they beyond truthful in many situations.As a result , we conclude that partly truthful,critical and consistent are three new directions in mechanism design.

1. *Introduction*

Mechanism design uses the tools of economics and game theory to design "rules of interaction" for economic transactions that will, in principle, yield some desired outcome. In the last few years this field has received much interest of researchers in computer science, especially with the Internet developing as a platform for communications and connections among enormous numbers of computers and humans. Arguably the most positive result in mechanism design is truthful and only one general truthful mechanisms : the generalized Vickrey-Clarke-Groves (VCG) mechanism. But VCG mechanism has one shortage:"the implementation of truthfulness is on the cost of decreasing the revenue of the mechanism." (e.g., Ning Chen and Hong Zhu. [1999]). We introduce three new characters of mechanism: partly truthful, critical and consistent, and introduce a new mechanism: X mechanism that satisfy the above three characters. Like VCG mechanism, X mechanism also generalizes from Vickery Auction and is consistent with Vickery auction in many ways, but the extended methods of X mechanism and VCG mechanism is different. This paper will demonstrate that X mechanism better than VCG mechanism in optimizing utility of mechanism , which is the original intention of utility of mechanism. So partly truthful , critical and consistent are at least as important as truthful in mechanism design, and they beyond truthful in many situations.As a result , we think partly truthful ,critical and consistent are are three New directions in mechanism design.

---

*Department of Computer Science, Nanjing University of Aeronautics and Astronautics, Nanjing 210016, P.R. China.

Mechanism design, a classic economic concept [Osborne et al. 1994], [ Papadimitriou, C.H. 2001], deals with algorithmic design problems involving in such human factors, as in task allocation problems (see Feigebaum, J. et al. [2000] and Noam Nisan and Amir Ronen. [1999] and W.E. Walsh et al. [1998], for example), communication networks (see Joan Feigenbaum et al. [2000] and Elias Koutsoupias et al. [1999], for example), multi-agent systems [Jeffrey S. Rosenschein and Gilad Zlotkin 1994], shortest paths problems [Noam Nisan and Amir Ronen 1998] and so on. Intuitively, mechanism design can be described as the design of protocols that realize a given target (objective function) under the assumption that the participants (agents) are all self-interested and rational ones who aim to optimize their own goals (ref Ning Chen and Hong Zhu [2004]). The main motivation of this field is micro-economic and the tools are game-theoretic.

The major achievement of mechanism design theory is VCG mechanism by Vickrey [1961], Clarke [1971], and Groves [1973] . It is the only one general truthful(also called incentive compatible, which means that the agents' interests are best served by behaving truthfully) technique known so far. It essentially provides a solution for any utilitarian problem (except for the possible problem that there might be dominant strategies other than truth-telling). It generalizes from Vickery Auction (named after Vickrey, who is widely recognized as the founder of auction theory and received the Economics Nobel Memorial prize in 1996.,which can motivate competing suppliers to reveal their true prices. It has been discussed, e.g., in article written by Green, J., Laffont, J.J. [1997], extensively. Nisan and Ronen [10, 19] studied this problem from the view of algorithmic design.

But when VCG mechanism is applied to complex mechanism design problems such as combinatorial auctions [Nisan, N 2000], shortest paths problems by Noam Nisan and Amir Ronen [1999], one shortcoming emerge: It excessively considers every agent's utility and ignores the mechanism's profit, to some extent, its total cost will be more larger than the traditional First Price Auction's in some case even if every agent raises their bidden (we will give an example in section 3).

To solve the shortcoming mentioned above, we introduce three new characters of mechanism: partly truthful, critical and consistent, and introduce a new mechanism: X mechanism that satisfy the above three characters. In fact, both X mechanism and VCG mechanism extend the Vickrey auction, but their extension methods are different. Later we will demonstrate that partly truthful, critical and consistent are three new directions in mechanism design, which are at least as important as truthful and can beyond truthful in many aspects.

## 2. *Preliminaries*

In this section we propose a formal model that assumes the participants all act according to their own self-interest. We adopt a rationality-based approach, using notions from game theory and micro-economics and particularly from the field of mechanism design. We assume that each participant has a well-defined utility function that represents its preference over the possible outputs of the algorithm, and we assume that participants rationally act to optimize their utility. We also assume that participants are independent. We term such rational and selfish participants agents. The solutions we consider contain both an algorithmic ingredient (obtaining the intended results) and a payment ingredient that motivates the agents. We term such a solution mechanism(references Noam Nisan and Amir Ronen [1999]).

Intuitively, a mechanism design problem has two components: the usual algorithmic output specification and descriptions of what the participating agents want, formally given as utility functions over the set of possible outputs.

**Definition** 2.1. *A mechanism design problem is given by an output specification and by a set of agents' utilities. Specifically:*
(1) There are n agents, each agent i has available to it some private input $t^i \in T^i$(termed its type). Anything else in this scenario is public knowledge.
(2) The output specification maps to each type vector $t = (t^1, ...t^n)$ a set of allowed outputs $o \in O$, where $O$ is a finite set of outcomes.
(3) Each agent $i$'s preferences are given by a real valued function: $v^i(t^i, o)$, called its valuation. This is a quantification of its value from the output $o$, when its type is $t^i$, in terms of some common units of this currency, then its utility will be $u^i = p^i + v^i(t^i, o)$. This utility is what the agent aims to optimize. (ref Noam Nisan and Amir Ronen [1999])

*Remark* 2.1. In some sense, we can approximately take $t^i$ as agent $i$'s cost price, and take $o$ as a outcome to state which agent is selected, $v^i(t^i, o)$ is the price agent $i$ bids, $p^i$ stands for the payment mechanism give to agent $i$ and take $u^i$ as the agent $i$'s profit.

**Definition** 2.2. *A (direct revelation) mechanism is a pair $m = (o, p)$ such that:*
(1) The output function o accepts as input a vector $V = (v^1, ..., v^n)$ of declared valuation functions and returns an output $o(V) \in O$(where $O$ is a finite set of outcomes).
(2) The payment function $p(V) = (p^1(V), ..., p^n(V))$ returns a real vector. This is the payment handed by the mechanism to each of the agents (e.g. if $p^i = -2$ then the agent pays two units of currency to the mechanism). (ref Ning Chen and Hong Zhu [2004])

*Remark* 2.2. In the above definition, the declaration $v^i$ of agent $i$ is not necessarily equal to $t^i$, since he may not tell the truth to the mechanism. His outcome is determined in terms of how to maximize his utility value $u^i = p^i + v^i(t^i, o)$ where $o$ is the outcome of the mechanism

*Remark* 2.3 We may define various outcome functions in different models. In utilitarian problems, the outcome function of the mechanism is to select an optimal outcome $o \in O$ that maximizes the objective function $g(o, v) = v^i(t^i, o)$, which is termed as social utility.

*Notation* 2.1: $(a^1, ...a^{i-1}, a^{i+1}...a^n)$ is denoted by $a^{-i}$. $(a^i, a^{-i})$ will denote the tuple $(a^1, ...a^n)$.

*Remark* 2.4. In a direct revelation mechanism, the participants are simply asked to reveal their types to the mechanism. Based on these declarations the mechanism then computes the output o and the payment $p^i$ for each of the agents. As they may lie to the mechanism, it should be carefully designed such that it will be for the benefit of each agent to reveal her true type to the mechanism.

**Definition 2.3** A mechanism is called truthful if truth-telling is a dominant outcome. I.e. for every agent $i$ of type $t^i$ and for every type declaration $t^{-i}$ for the other agents, the agent's utility is maximized when she declares her real valuation function $t^i$. (Ref Noam Nisan and Amir Ronen [2000])

*Remark* 2.5. One desirable property of mechanisms is that the utility of a truthful agent is always non-negative. This is often called participation constraints (ref Nisan, N's [1999]).

*Remark* 2.6. A outcome is dominant if, regardless of what any other players do, the outcome earns a player a larger payoff than any other. In truthful-telling mechanism, Truth-telling is a dominant outcome.

Next we will talk about Vickrey Auction and analyze its properties. The well-known Vickrey Auction is a sealed-bid second price auction in which each participant simultaneously submits bids. The auctioneer discloses the identity of the highest agent who is declared as the winner. The price paid, however, is equal to the second-highest bidding. This format is named after William Vickrey who first described it and pointed out that agents have a dominant outcome to bid their true values.

The Vickrey Auction have three significant characters:
(1)Truthful.
(2)Critical: There will be two possible choice if the mechanism pays a little more. If mechanism increase its utility a little, there are only one possible outcome;In other words, the auctioneer needn't pay more money than second-highest price,otherwise he can choose agent who bids the second-highest price.
(3)Consistent: The outcome of mechanism can maximize the profit of both mechanism and that of auctioneers at the same time.

**Definition** 2.4. A mechanism $m = (o, p)$ belongs to VCG family if:
(1) $o(w) \in \arg\max_{o \in O} \sum_{i=1}^{n} v^i(t^i, o)$.
(2) The payment is calculated according to the VCG formula $p^i(V) = v^j(o(V)) + h^i(V^{-i})$,where $h^i$ is an arbitrary function of $V^{-i}$.

THEOREM 1 ([3]) A VCG mechanism is truthful.

Note that because of truthfulness, every agent indeed reports his true type $t^i$ in VCG mechanism, i.e., $v^i = t^i$.

Many algorithmic mechanism design problems can be solved using the VCG mechanism, for example task allocation problems([Feigebaum, J. et al. 2000]) and combinatorial auctions studied

by Nisan, N. [2000]. In this paper, we will discuss mechanism design mainly about the Shortest Path Problem .

The description of Shortest Path Problem is as follows: We are given a directed graph $G$ with two distinguished nodes $s$ and $t$. Each edge $e$ in the graph is owned by a self interested agent i, and I(e) denotes the set of edges belonging to agent i. The actual cost of routing an object along an edge e (denoted by $v_e$) is privately known to its owner. The goal of the center is to acquire an path from $s$ to $t$ and pay as little money as possible. This goal is not shared by the participating agents. Each agent selfishly wants to maximize her own profit which is the difference between her payment $p^i$ and her actual costs, that is, $u^i = p^i - \sum_{e \in P \cap I(e)} v_e$, where P denotes the chosen path. We assume that no agent owns a cut in the graph.

When all agents honestly report their costs in VCG mechanism, the cheapest path is chosen: the output is obtained by a simple shortest path calculation. The payment given to agent $i$ is 0 if $i$ is not in the shortest path and $p^i = (d_{i=\infty} - d_{i=0})$ if it is , here $d_{i=\infty}$ is the length of the shortest path which does not contain $i$ , and $d_{i=0}$ is the length of the shortest path when the cost of $i$ is assumed to be zero. Notice that the shortest path is indeed a minimization of the total cost.

In the Figure 1, Suppose sender X is looking for a path to Y for transmiting something. There are several paths from X to the Y. Each edge on the path stands for an selfish agent. The number beside every agent stands for its cost $v^i$. We can find the cheapest path is $X \to A \to B \to C \to D \to E \to F \to Y$. According to VCG mechanism: $p^A = d_{A=\infty} - d_{A=0} = (v^J + v^P + v^E + v^F) - (v^B + v^C + v^D + v^E + v^F) = (4+4+1+1)$

Example 1:

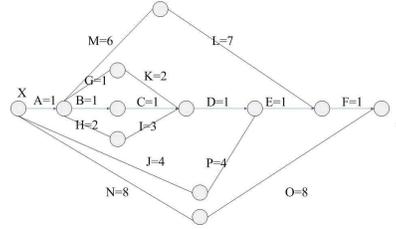

$Figure 1$

$-(1+1+1+1+1) = 5$
$p^B = d_{B=\infty} - d_{B=0} = (v^A + v^G + v^K + v^D + v^E + v^F) - (v^A + v^C + v^D + v^E + v^F) = (1+1+2+1+1+1) - (1+1+1+1+1+1) = 1$
$p^C = d_{C=\infty} - d_{C=0} = 1$ , $p^D = d_{D=\infty} - d_{D=0} = 5$ , $P^E = 10$ , $P^F = 11$

The total cost of sending a single message along this cheapest path is $p^A + p^B + p^C + p^D + p^E + p^F = 33$, which is twice more than the most expensive path $(v^N + v^O = 16)$.

We analyze disadvantages of the VCG mechanism by example 1:
(1) For making up of the cheapest path, edge B is as important as edge C, because without either of them the cheapest path will change to $X \to A \to G \to K \to D \to E \to F \to Y$. They should share the payment of contributions to make up the cheapest path. But in the VCG mechanism, the payment of each edge is $p^i = d_{i=\infty} - d_{i=0}$. It gives the total payment to every edge and assuming other edge's profit is 0.This will makes the sum of payment be large.
(2) It can be observed that edge B and edge C make different contributions from edge A. The shortest path when delete the edge B or C is Because of the (1)(2)above, they should belong to different groups and share the payment of their corresponding groups. Edge B and edge C should belong to the same group, edge A should belongs to another group.

We have analyzed the Vickrey auction has a significant merit: Critical. But by Example 1, we can see that the VCG mechanism has not this significant merit. With the payment 33 currency settled by VCG mechanism , all the agents whose profit above zero will accept the payment, so the mechanism can choose any path which the total bidden less than 33. There are six pathes from X to Y can be selected by sender X:
$X \to A \to B \to C \to D \to E \to F \to Y$ , $X \to A \to G \to K \to D \to E \to F \to Y$ , $X \to A \to H \to I \to D \to E \to F \to Y$ , $X \to J \to P \to E \to F \to Y$ , $X \to A \to M \to L \to F \to Y$ , $X \to N \to O \to Y$.

We can also observe that in the above example, if edge A raises his price, the total cost of mechanism will go down in the VCG mechanism. It shows that the payment by VCG mechanism is not the optimal for the auctioneer. In above example , If the bidding of A is rise to 4 ,then the payment of X by VCG will be computed as follows:

The cheapest path from x to y is $X \to A \to B \to C \to D \to E \to F \to Y$. $p^A = d_{A=\infty} - d_{A=0} = (v^J + v^P + v^E + v^F) - (v^B + v^C + v^D + v^E + v^F) = (4+4+1+1) - (1+1+1+1+1) = 5$, $p^B = 1$, $p^C = 1$, $p^D = 2$, $P^E = 10$, $P^F 7$ ,The total cost of sending a single message along this cheapest path is $p^A + p^B + p^C + p^D + p^E + p^F = 26$. So , A bid 4 is more favorable than A bid 1 for the sender x .

*Remark* 2.7. In VCG Mechanism, the truth-telling outcome is more profitable for agent than for the mechanism. The implementation of truthfulness is on the cost of decreasing the revenue of the mechanism. Thus we should find a weaker feasible notion of truthfulness and new mechanisms to balance the profits of Mechanism and agent

## 3. A new mechanism: X mechanism

For simplicity, we discuss X mechanism mainly about the shortest path problems in this section.

*Notation* 3.1: In following , $o_1$ denote the best outcome in $O$, $o_2$ stands for the second best outcome ... $o_s$ stands for the s-th best outcome.

*Notation* 3.2: The cost of each outcome is denoted by $s_j = \sum_{i \in o_j} v^i$

**Definition** 3.1 If agent $i$ is in $o_1, o_2, \ldots$ and until $o_q$ simultaneously, but not in $o_{(q+1)}$ , then agent $i$ belongs to the q-th group , denoted by $k^i = q$ or $i \in g_q$.

For example, in the Shortest Path problem , for every agent $i$ , the agent's $k^i = s$ when agent I's $v^i$ is in the cheapest path, in the second cheapest path , ... in the s'th cheapest path $o_s$ simultaneously and not in (s+1)th cheapest path $o_{s+1}$.

*Remark* 3.1 In the **Definition** 3.1 we only classify the agents in $o_1$ to groups and other agents are not classified to groups.

**Definition** 3.2 (X mechanism)
A mechanism m = (o, p) belongs to X mechanism if:
(1) $o(w) \in \arg max_{o \in O} \sum_{i=1}^{n} v^i(t^i, o)$
according to the *Notation* 3.1, it is $o_1$.
(2)According to the **Definition** 3.1, Agents in the outcome of o(w) are divided into some groups : $(g_1, g_2, \ldots, g_h)$.(suppose there are h groups in the shortest path)
(3) The total profit of all agents in group $g_i$ is denoted by $Q_{g_i}$ , it can be computed as follows: $Q_{g_i} = s_{((g_i)+1)} - s_{(g_{(i-1)}+1)}$
(4)The pure profit of agent $j \in g_i$ is calculated according to the formula $H(v^j, V^{(g_i)}, Q_{g_i})$, where $V^{(g_i)}$ denote all bidding of agents which is in group $g_i$ , H is an arbitrary function and $H > 0$. The payment to agent j is $p^j = v^j + H(v^j, V^{(g_i)}, Q_{g_i}) > v^i$.

*Remark*3.2: Here the total pure profit of group i stands for the difference between the total payment and the total bid of agents in group i.

*Remark*3.3: Suppose agent j$\in g_i$ and the number of agents in group $g_i$ is $m(m \geq 1)$. There are many possible formula for H, some of them be listed as follows:
3.2.1 $H(v^j, V^{(g_i)}, Q_{g_i}) = Q_{g_i}/m$
3.2.2 Each agent in group i gets his profit according to the reverse sequence of his input in all bidding of agents which is in group $g_i$. Suppose the bid of agent w is the j-th big in group i and the bid of agent y is (m-j+1)-th big in group i, The formula is:
$P^w = v^w + (v^y/(\sum_{k \in g_i} v^k)) \times Q_{g_i}$

For example,there are three agents a,b,c(i.e. m=3) in group $g_i$, and $v^a = 10$, $v^b = 20$, $v^c = 30$, then $P^a = (10 + (30/(10+20+30)) \times Q_{g_i})$, $P^b = (20 + (20/(10+20+30)) * Q_{g_i})$, $P^c = (30 + (10/(10+20+30)) \times Q_{g_i})$.

3.2.3 The mechanism firstly distributes a minimum of profit $\Delta$ to every agent of group z, then distributes the rest of profits to the agent i(suppose the bidding of agent i is lowest in group z) until equal to the bidding of agent j (suppose the bidding of agent j is second lowest in group z), and then distribute the rest of profits to agent i and agent j until their payment equal the distributing process until the payment of all agents are the same or the profits are all distributed. The exact computation is as follows

1. compute j which satisfies the following formula($\triangle$ stands for the minimum profit, and suppose the smallest bid in $g^i$ which is bigger than $v^j$ is $v^{j+1}$)

$$m \times \triangle + v^j \times j + \sum_{k \in g^z \text{ and } v^k > v^j} v^k <= Q + \sum_{k \in g^z} v^k < m \times \triangle + v^{j+1} \times (j+1) + \sum_{k \in g^z \text{ and } v^k > v^{j+1}} v^k$$

2. $P^k = (Q - m \times \triangle + v_j \times j + \sum_{k \in g^z \text{ and } v^k > v^j} v^k)/j + v^j \quad (k \in g^z \text{ and } v^k <= v^j)$

$P^k = v^k + \triangle \quad (k \in g^z \text{ and } v^k > v^j)$

For example, there are three agents a,b,c in group $s_i$(i.e. m=3), and $v^a = 10$, $v^b = 20$, $v^c = 30$, $\triangle = 1$, $Q_{s_i} = 15$, then the process of distribute the payment of group $V_j$ can be denote by $(v^a, v^b, v^c) = (10, 20, 30) \rightarrow (11, 21, 31) \rightarrow (21, 21, 31) \rightarrow (22, 22, 31) = (p^a, p^b, p^c)$.

*Remark*3.4: The function H in *Remark* 3.2.1 and the function H in *Remark* 3.2.2 represent two extreme distributing strategies and The function H in *Remark* 3.2.3 combines them. The character of function H in *Remark* 3.2.1 is distributing the payment as averagely as possible. The character of function H In *Remark* 3.2.2 is letting the following condition as possible as be satisfied: The lower the bidding of agent is, the higher its pure profit is. Thus H in *Remark* 3.2.2 and in *Remark* 3.2.3 is better than H in *Remark* 3.2.1 in encouraging bidders report their type.

In addition, we also can use the compounding-form of the three above. For example, firstly we use the third method, then the spare profits are distributed by the first or second method. The objective of various forms for function H is to make every agent tell his valuation truthfully, (because H can make the agent whose valuation is lower get more pure profit.) to make the Critical mechanism is truthful. But it is a pity that no matter which form is chosen, for individual agent, it is not truthful. It is just close to truthful. We will use combinatorial mechanism to solve this problem later.

*Remark*3.5: In X mechanism, we consider the agents of group s as a single entity, and all agents in this group share the pure profit of the group $Q_s$. We denote the total pure profit of group s by $Q_s$, and all agents in this group share $Q_s$ according to function H in *Remark* 3.3

Now we apply X mechanism to the Example 1 (according to figure 1)as follows:
the shortest path $o_1$ is:
$X \rightarrow A \rightarrow B \rightarrow C \rightarrow D \rightarrow E \rightarrow F \rightarrow Y$, $s_1 = 6$, the second shortest path $o_2$ is : $X \rightarrow A \rightarrow G \rightarrow K \rightarrow D \rightarrow E \rightarrow F \rightarrow Y$, $s_2 = 7$, the third shortest path $o_3$ is : $X \rightarrow A \rightarrow H \rightarrow I \rightarrow D \rightarrow E \rightarrow F \rightarrow Y$, $s_3 = 9$, the fourth shortest path $o_4$ is : $X \rightarrow J \rightarrow P \rightarrow E \rightarrow F \rightarrow Y$, $s_4 = 10$, the fifth shortest path $o_5$ is : $X \rightarrow A \rightarrow M \rightarrow L \rightarrow F \rightarrow Y$, $s_5 = 15$, the sixth shortest path $o_6$ is: $X \rightarrow N \rightarrow O \rightarrow Y$. $s_6 = 15$

In figure 1, 1-group contain edge B and C (denoted by $g_1$={B,C}, for they belong to $o_{(1)}$ and not belong to $o_{(2)}$; edge A and D belong to the 2-group, for they belong to $o_{(1)}$ ,$o_{(2)}$, $o_{(3)}$ ,and not belong to $o_{(4)}$ ; edge E belongs to the 4-group; edge F belongs to the 5-group.
The profit of 1-group is $Q_1 = s_2 - s_1 = 7 - 6 = 1$, The profit of 2-group is $Q_2 = s_4 - s_2 = 10 - 7 = 3$, The profit of 3-group is $Q_3 = s_5 - s_4 = 15 - 10 = 5$, The profit of 4-group is $Q_4 = Q_4 = s_6 - s_5 = 16 - 15 = 1$,
If the function H in 3.2.1 is adopted, then $P^B = v^B + H(Q_1) = 1 + 0.5 = 1.5$, $P^C = v^C + H(Q_1) = 1 + 0.5 = 1.5$, $P^A = v^A + H(Q_3) = 1 + 1.5 = 2.5$, $P^D = v^D + H(Q_3) = 1 + 1.5 = 2.5$, $P^E = v^E + H(Q_4) = 1 + 5 = 6$, $P^F = v^F + H(Q_5) = 1 + 1 = 2$.

*Remark*3.6 $Q(i) = Q(g_i)$ be equal to$(s_{((g_i)+1)} - s_{(g_{(i-1)}+1)})$ and need not be equal to $((s_{(i+1)}) - (s_i))$, For example in figure 1 ,$g_1 = 1, g_2 = 3, Q_2 = (s_{g_i+1} - s_{g_{(i-1)}+1}) = (s_4 - s_2) \neq (s_4 - s_3)$.

*Remark*3.7 X mechanism and VCG mechanism both extend the Vickrey auction. But their extend manner is different. VCG extends Vickery auction by ways of considering agent's profit and preserves truthful characteristic; X mechanism extends Vickery auction by ways of considering mechanism's profit and remains the critical characteristic.

*Remark*3.8 In X mechanism, the total payment is
$C = s_{v_1} + \sum Q_{v_i} = s_{v_1} + (s_{v_2} - s_{v_1}) + (s_{v_3} - s_{v_2}) + \ldots (s_{v_h} - s_{v_{(h-1)}}) + (s_{v_{(h+1)}} - s_{v_h}) = s_{v_{(h+1)}}$.
(Suppose $v_h$ is the index number of maximum group in the cheapest path)

Remark 3.9 When there is only one member in a group the member will choose the bidden the same as he is under truthful mechanism. Because raising price will increase risk and will not increase profit.

## 4. Three new characters of mechanism

In this section , we will discuss three new characters of mechanism : (strongly)partly truthful , (Strongly) Critical ,(Strongly ,Partially ,Impossible)Consistent .

**Definition** 4.1 Any mechanism which satisfies the following three conditions is called partly truthful mechanism.
(1)When agent bid his type , its probability of being selected is the highest.
(2)The higher the agent's bidding is, the lower the probability of his being selected is.
(3)The utility of the agent is more than zero when he is selected.

*Remark* 4.1 The difference between partly truthful mechanisms and truthful mechanisms is that in truthful mechanism the highest probability of agent being selected and the highest utility of agent will attain at the same time when agent bids his type. But in partly truthful mechanism, when the agent bids his type, only his probability of being selected is the highest. So truthful mechanisms are the subsets of partly truthful mechanisms.

*Remark* 4.2 The risk-avert agent or rational agent will choose to bid true price in partly truthful mechanisms.

**Definition 4.2:** A mechanism is called strongly partly truthful mechanism , if agent i raise his bidden in a certain range, the payment to it will be unchanged, but the probility of its not been seclected will be enlarged at the same time ; if it raise his bidden above this certain range , then the payment to it will be enlarged under the condition that it is also be selected all the same , and the probility of its not been seclected will be enlarged at the same time.

*THEOREM* 4.1 X mechanism is truthful to the all agents selected by the mechanism. The total payment of X mechanism is $C = s_{v_{(h+1)}}$(the meaning of h is the same as that in *Conclusion* 4.1). It is not related to the biddings of agents which are in cheapest path. So it is truthful. If there is only one edge on the Shortest Path, X mechanism degenerate to Vickrey auction.

*THEOREM* 4.2 X mechanism is partly truthful .
Prove: $p^i = v^i + H(v^j, V^{(g_i)}, Q_{g_i})$ and $H(v^j, V^{(g_i)}, Q_{g_i}) > 0$ , the profit of agent i is certainly above zero and its probability of being selected is the highest when it bid its type . So X mechanism is partly truthful .

*THEOREM* 4.3 In x mechanism , it is truthful for each group $g_i$.

**PROOF** We assumed when agents of group $g_i$ changing their valuation ,the valuation of other agents remain unchanged and all the original outcome$(o_1, o_2, \ldots)$ don't change. Then the total payment to group $g'_i s$ $P(g_i) = v^{g_i} + Q_{g_i} = v^{g_i} + s_{((g_i)+1)} - s_{(g_{(i-1)}+1)}$ is unchanged because when $v^{g_i}$ increase some quantity , the $Q_{g_i}$ will decrease the same quantity. In a word , $P(g_i)$ is unchanged even when $v^{g_i}$ increase in some condition , so it is truthful for each group $g_i$.

*Remark* 4.3 In the Shortest Path problem , mechanism can increase its utility by decreasing its payment or decrease its utility by increasing its payment.

**Definition** 4.3 A mechanism belongs to Critical mechanism if it satisfies the following three conditions:
(1)It is partly truthful mechanism.
(2) If mechanism increase its utility a little, there are only one possible outcome;
(3)If mechanism decrease its utility a little, there are at lease two possible outcome.

*THEOREM* 4.4 Vickrey auction is critical.

**Definition** 4.4 A mechanism belongs to Strongly Critical mechanism if it satisfies the following two conditions:
(1)It is Critical mechanism.
(2)The payment given to any group which been selected in the shortest path always is a critical value. In other words, If the payment to any group which been selected in the shortest path increase a little, then the mechanism can choose another path to substitute this shortest path by this increased payment.

*THEOREM* 4.5 X mechanism is a Strongly Critical mechanism .

*THEOREM* 4.6 VCG mechanism is not a Critical mechanism .

In following definitions , we assume the number of the agents is n. The bid of agent is denoted by $v^i$ , the utility of agent i is denoted by $u^i$. $(v^1, v^2, \ldots, v^{i-1}, v^{i+1}, \ldots, v^n)$ is denoted by $v^{-i}$ , and the possibility of agent i being chosen when its bidding is $v^i$ is denoted by $pr(v^i)$.

**Definition** 4.5 $OBS^i(A)=\{v^i|v^i$ can make $u^i$ maximum in some $v^{-i}$ at mechanism A$\}$

In some $v^{-i}$ at mechanism A , the Optimal bidding set of agent i $(OBS^i(A))$ is the set that contains only every value of $v^i$ which possibly make $u^i$ maximum in this situation.

**Definition** 4.6 At mechanism A, the Optimal bidding set of all agent's is denoted by OAB(A)=$\{(v^1,v^2,\ldots,v^n)|v^1\in OBS^1, v^2\in OBS^2,\ldots,v^n\in OBS^n,\}$.

**Definition** 4.7 AES(A)=$\{(v^1,v^2,\ldots,v^n)|(v^1,v^2,\ldots,v^n)$ can make the profit of mechanism A maximum$\}$

**Definition** 4.8 IOA(A)= $OAB(A)\cap AES(A)$ . The Intersection of OAB(A) and AES(A) is denoted by IOA(A).

**Definition** 4.9 Mechanism A is Consistent if $IOA(A)\neq\emptyset$ in any situations.

**Definition** 4.10 Mechanism A is Strongly Consistent if IOA(A)=OAB(A) or IOA(A)= AES(A) in any situations.

**Definition** 4.11 Mechanism A is Partially Consistent if $IOA(A)=\emptyset$ in some cases and $IOA(A)\neq\emptyset$ in other cases.

**Definition** 4.12 Mechanism A is Impossible Consistent if $IOA(A)=\emptyset$ in any situations.

Suppose the agents and the mechanism tries to maximum the mathematical expectation of their utility , we will give some examples of above definitions: X Mechanism is a Consistent mechanism ,the First Price Auction is an Impossible Consistent Mechanism, the Vichrey Auction is a Strongly Consistent mechanism, the VCG mechanism is a Partially Consistent mechanism. Detailed analysis as follows:

(1) The First Price Auction is an Impossible Consistent mechanism

In First Price Auction mechanism, the agent has two strategy of bidding , the first is bidding his type $t^i$, the second is decreasing his bidding to $v^i$ ($0<v^i<t^i$) for gaining more profit ($u^i=t^i-v^i$) but at the same time his probability of be chosen decreases. In first strategy ,$v^i=t^i$ , $u^i(v^i=t^i)=t^i-v^i=0$, the expectation of his utility is $u^i\times pr(t^i)+0\times(1-pr(t^i))=0$ . In second strategy, $0<v^i<t^i$, $u^i(v^i<t^i)=t^i-v^i>0$ and the possibility of being chosen $0<pr(v^i)<pr(t^i)$ , the expectation of his utility is $u^i(v^i<t^i)\times pr(v^i)+0\times(1-pr(v^i))>0$ .

According to expectation of utility , agent chooses the second strategy of bidding , then $OBS^i$ (First Price Auction)=$\{v^i|0<v^i<t^i\}$. So OAB(First Price Auction )=$\{(v^1,v^2,\ldots,v^n)|0<v^i<t^i, 0<v^2<t^2,\ldots,0<v^n<t^n\}$.

Obviously, the auctioneer's utility is decided by the the highest bidding, so AES(First Price Auction )=$\{(v^1,v^2,\ldots,v^n)|$ the bidder of highest bidding bid his type$\}$

IOA(First Price Auction)=OAB(First Price Auction )$\cap$ AES(First Price Auction ) $=\emptyset$.

So, First Price Auction is an impossible Consistent mechanism.

(2) Vichrey Auction is a Strongly Consistent Mechanism.

Because Vichrey Auction is truthful , OAB(Vichrey Auction)=$\{(t^1,t^2,\ldots,t^n)\}$. AES(Vichrey Auction)= $\{(v^1,v^2,\ldots,v^n)|$ the bidder of second highest bidding bid his type$\}$,

IOA(Vichrey Auction)=OAB(Vichrey Auction)$\cap$ AES(Vichrey Auction) =$\{(t^1,t^2,\ldots,t^n)\}$= OAB(Vichrey Auction).

So Vichrey Auction is a Strongly Consistent Mechanism .

(3) VCG mechanism is a Partially Consistent Mechanism

Because VCG mechanism is truthful , OAB(VCG mechanism)=$\{(t^1,t^2,\ldots,t^n)\}$.

To explain the computation of AES(VCG mechanism) , we take two example of Shortest Path problem . In the following Graph , Suppose sender X is looking for a path to Y for transmiting something. There are several paths from X to the Y. Each edge on the path stands for an selfish agent. The number beside every agent stands for its type $t^i$.

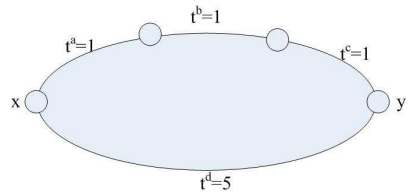

*Figure* 2

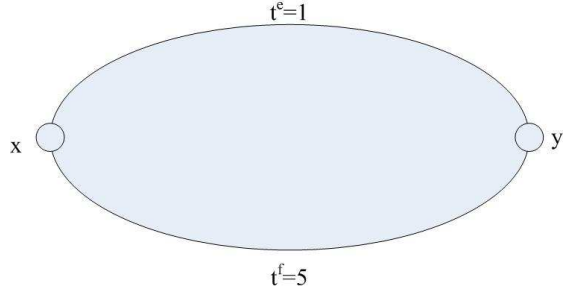

*Figure 3*

Suppose the minimum unit of currency is 1 and no two paths have the same length. In graph 2, AES(VCG mechanism)=$\{(v^a=2, v^b=1, v^c=1, v^d=5), (v^a=1, v^b=2, v^c=1, v^d=5), (v^a=1, v^b=1, v^c=2, v^d=5)\}$, OAB(VCG mechanism)=$\{(v^a=t^a=1, v^b=t^b=1, v^c=t^c=1, v^d=t^d=5)\}$, So IOA(VCG mechanism)=OAB(VCG mechanism)∩ AES(VCG mechanism)=∅

In graph 3, AES(VCG mechanism)=$\{(v^e=1, v^f=5), (v^e=2, v^f=5), (v^e=3, v^f=5), (v^e=4, v^f=5)\}$, OAB(VCG mechanism)=$\{(v^e=1, v^f=5)\}$, So IOA(VCG mechanism)=OAB(VCG mechanism)∩ AES(VCG mechanism)= $\{(v^e=1, v^f=5)\} \neq \emptyset$.

Thus, VCG mechanism is a Partially Consistent mechanism.

(4) X Mechanism is Consistent mechanism

Considering X Mechanism in Shortest Path problem, when agent i bid its type($v^i=t^i$), the possibility of its been selected is maximum, and its profit is bigger than 0; when the bidding of agent i is more than its type($v^i > t^i$), the possibility of its been selected is less than that of when it bid its type, and its profit is bigger than that of when it bid its type.

For the risk-avert agents and the agents who don't want to spend much of time in collecting information of other agents' bidding, they will choose bid their type, and their profit bigger than 0. In this situation, X mechanism can attain optimal results. For the agents who pursue high profits even if the risk of the possibility of being not selected become higher, they choose to raise their bidding which bigger than its type.

So $OBS^i$(X Mechanism)=$\{v^i | v^i \geq t^i\}$
OAB(X Mechanism)=$\{(v^1, v^2, \ldots, v^n) | v^1 \geq t^1, v^2 \geq t^2, \ldots, v^n \geq t^n\}$

Suppose the maximum number of group in shortest path is k in X Mechanism, then the total payment to the agents in shortest path is the sum of bidding of the agents in the (k+1)-th shortest path. So AES(X Mechanism)=$\{(v^1, v^2, \ldots, v^n) |$ the bidder of the (k+1)-th shortest path bid their types$\}$

IOA(X Mechanism)= OAB(X Mechanism)∩ AES(X Mechanism) =$\{(v^1, v^2, \ldots, v^n) | v^j = t^j$ for any bidder j in the (k+1)-th shortest path, and for any other agents i, $v^i \geq t^i\} \neq \emptyset$.

So X Mechanism is Consistent Mechanism.

## 5 Compare X mechanism and other mechanism

5.1 the total payment mechanism

If the agent is risk-avert, then every agent will report their type in X mechanism or VCG mechanism. In X mechanism can output with lower total cost than VCG mechanism.

If there is only one node on the cheapest path, VCG mechanism and X mechanism are the same with Vickery auction.

If the agent is not risk-averterhe may probably raise the price under our mechanism but he will not raise price under VCG mechanism. In this case, each mechanism is possibly better than other mechanism, according to the concrete situation of all agents' raising price.

For example, in above example 1, if the agents in the sixth shortest path(N and O) raise their bidding less than or equal to two times of their original bidding, then the total payment of X mechanism is less than that of VCG mechanism ; if they raise their bidding more than two times of their original bidding, then the total payment of X mechanism is more than that of VCG mechanism.

### 5.3 inherit and change

Vickery auction is the specific example of X mechanismin which Every class has only one edge. Characteristics of truthful mechanism is the money given to the first selected bidders is less than the second selected bidders when the first selected bidders all quit. The extend manner is different. VCG extends Vickery auction by ways of considering agent's profit and preserves truthful characteristic; X mechanism extends Vickery auction by ways of considering mechanism's profit and remains the Balanced characteristic. And when every group has only one member k-Group and VCG is the same.

So X mechanism and VCG mechanism both extend the Vickrey auction. Now we apply X mechanism to the Example 1,as follows: Shortest Path: XABCDY Second Shortest Path : XABGHY In our mechanism, A and B are 3-group points. C and D are 1-group points. The profit of 1-group points is Q1,Q1 = g2 - g1 = 12 - 6 = 6. The profit of 3-group points is Q3, Q3 = g4 - g3 = 25 - 14 = 11 . The pure profits of point A,B,C,D are PA,PB,PC,PD respectively. PA = wA + H( Q3 ), PB = wB + H( Q3 ), PC = wC + H( Q1 ), PD = wD + H( Q1 ),

Truthful mechanism only consider optimizing the $g(o,w) = v^i(t^i, o)$ , even if at the cost of $p(o,w) = \sum_{i=1}^{n} p^i$ sometimes. X mechanism not only take account of optimizing the $g(o,w)$, but also take account of minimize $p(o,w)$. It as possible as pays attention to both above object at the same time. In the situation that can not attain the two above object at the same time, X mechanism first considered minimizing $p(o,w)$.

X mechanism is a compromise of the First Price Auction and the Vichery Auction.

In the First Price AuctionAn auction in which the agent who submitted the highest bidden is awarded the object being sold and pays a price equal to the amount bid., the lower bidding price it is, the bigger chance that the agent will choose it. When the chance of the agent be selected is the biggest (bid his type), the profit it will get is zero.

In the Vichery Auction, auctioneer will give the second-highest price to the agent who submitted the highest price. The payment to the agent which submitted the highest price is not related to the price that he submitted. When the chance of the agent be selected is biggest (tell his type), the profit it will get is biggest.

In X mechanism , it is truthful for each group in the shortest path. If agent in one group raises his valuation his payment will increase. But the chance for him to be selected will decline. So when the chance for the agent to be selected is biggest (bid his type), X mechanism can make sure his profit is more than zero but not always the greatest as that of Vichery Auction (Except there is only one agent in the shortest path). What is more, the total cost of X mechanism ($p(W)$) is smaller than that in VCG mechanism under the conditions that the bid of all agent is unchanged.

There are an improvement to the First Price Auction. The payment to the agent who submitted the highest price be the diffence of its bidding and a constant $C$ .On the surface it works like X mechanism, the profit of every agent is bigger than zero when its chance be seclected is biggest. But in fact, in order to gain bigger chance, agents will adding $C$ to his original bidding . So when the chance of agents be seclected is biggest , (under the participation constraints ), the profit of agents is zero .

### 5.4 computing complexity

The computing complexity of X mechanism is the same as that of the VCG mechanism.

In X mechanism , the computing is just as same as that in VCG mechanism. In VCG mechanism, $p^i = d_{i=\infty} - d_{i=0}$, when compute $d_{i=\infty}$, the new shortest path should be computing. If $i$ belong to k-group, then when $v^{i=\infty}$ , the new shortest path is the $(k+1)$-shortest path. In X mechanism, when computing the group of the agent $i$ in the cheapest path, let $v^{i=\infty}$ and computing the new shortest path . If $i$ belong to $k$-group, then the new shortest path is $(k+1)$-shortest path. Both of the two above mechanism need to calculate the shortest path and new shortest path when each agent in the shortest path is removed.

Suppose the number of agents in the Shortest path is n and there are k groups in the Shortest Path , there are only two things that not compute in VCG mechanism should be computed in X mechanism : the one is to calculate the length cost of the $(k + 1)$-shortest path, the other is put the edges that is removed will lead to the same Shortest Path in one group and sorting the length of new Shortest Path which is formed when some edge is removed.

There are some algorithms in the shortest path problem which can lead to $O(V + E)$ or $O(V^2)$

time complexity [,, $k$-th shortest path,jacm] , where V is the number of the vertex, E is the number of the edges.There are some algorithms in the sorting problem whose time complexity are $m(lnm)$ , therein m is the number of elements. So the time complexity of VCG is $O(n*(V+E))$, and the time complexity of X mechanism is $O(n*(V+E)+(V+E)+n(ln(n))=O(n*(V+E))$ (because $V > n$) So the time complexity of X algorithmic is as the same as that of VCG mechanism.

## 6. Tradeoff of Truthful mechanism and X mechanism

We take VCG mechanism as an example to introduce truthful mechanism below. Through the above discussion we know that for each group the X mechanism is truthful. But for each agent, it's partly truthful. Therefore, each agent will rise up the price in order to increase his profit. But we know that the VCG mechanism is truthful. So we can combine the X mechanism with VCG mechanism . For example, we list three possible form of combinatorial mechanism as follows:

### 6.1 Tradeoff mechanism 1

**Definition** 6.1.(combinatorial mechanism) A mechanism $m = (o, p)$ belongs to combinatorial mechanism if:

(1) $C$ is given in advance, $0 =< C <= 1$

(2) the outcome of VCG mechanism is termed as $o_I(w)$, and the outcome of X mechanism is termed as $o_{II}(w)$.

(3) if (Payment($o_I(w)$) - Payment($o_{II}(w)$))/ Payment($o_I(w)$) > $C$, then the payment method of X mechanism is adopted, otherwise the payment method of VCG mechanism is adopted.

**Remark 6.1**. Payment($o_I(w)$) $\geq$ Payment ($o_{II}(w)$). (Proved in above )

### 6.2 Tradeoff mechanism 2

every agent of the $j$-th group gain the same payment, which is $(s_{j+1} - s_j)$ .Obviously , this is more truthful than the X mechanism and less than VCG. In fact , this mechanism is strongly partly truthful.

With the bidding increasing , the profit will increasing non-linearly. Suppose the original bidding of agent $i$ is $v^i$ , and it is in group $k$.

When $\Delta v^i \leq s_{k+1} - s_k$,
$P^i = (s_{k+1} - s_k) + v^i$
$= (s_{k+1} - (s_k + \Delta v^i)) + (v^i + \Delta v^i)$
$= (s_{k+1} - s_k) + v^i$

if $s_{k+1} - s_{k-1} \geq \Delta v^i > s_{k+1} - s_k$,
$P^i = (s_{k+1} - s_{k-1}) + v^i$
$= (s_{k+1} - (s_{k-1} + \Delta v^i)) + (v^i + \Delta v^i)$
$= (s_{k+1} - s_{k-1}) + v^i$

$s_{k+1} - s_{k-2} \geq \Delta v^i > s_{k+1} - s_k$,
$P^i = (s_{k+1} - s_{k-2}) + v^i$

$\vdots$

$s_{k+1} - s_1 \geq \Delta v^i > s_{k+1} - s_2$,
$P^i = (s_{k+1} - s_1) + v^i$

$\Delta v^i > s_{k+1} - s_1$,
$P^i = 0$

### 6.3 Tradeoff mechanism 3 ( )

Combinatorial mechanism in nature is that it would rather pay more in a certain range to motivate agents to tell their types truthfullyto increase the truthfulness of mechanism at the cost of critical. It can make the X mechanism near to truthful.

Compare the four mechanisms:

1.X mechanism: All the agents of the the $k$-th group share the difference between the total bidding of the $(k + 1)$-th group and that of the $k$-th group.

2. Tradeoff mechanism 2:Every agent of the the $k$-th group obtain the same profit respectively, which is the difference between the total bidding of the $(k+1)$-th group and that of the $k$-th group.

3. Tradeoff mechanism 3:All the agents of the the $k$-th group share the difference between the total bidding of the $(k + 1)$-th group and that of first group.

4.VCG mechanism: Every agent of the the $k$-th group obtain the same profit respectively, which

|  | The First Price Auction | | Vickery Auction | | VCG mechanism | | NW equilibrium mechanism | |
|---|---|---|---|---|---|---|---|---|
|  | telling the truth | Raising valuation | Telling the truth | raising valuation | telling the truth | raising valuation | telling the truth | raising valuation |
| Probability of choosen | maximum | ↓ | maximum | ↓ | Maximum | ↓ | maximum | ↓ |
| Profit when choosen | 0 | ↑ | $S_0$ | $S_0$(not change) | S | S(not change) | $0<q<s$ | $r<q'<s$ |
| Profit when not choosen | 0 | 0 | 0 | 0 | 0 | 0 | 0 | 0 |

**Future works**

In the Vichery Auction, the payment to the agent that bid the highest bidding (suppose this agent is agent 1) is only related to the second highest bidding $v^2$, so $p^1 = F(v^2)$.

In the First Price Auction, the payment to the highest bidder 1 is only related to the the bidding of itself. so $p^1 = F(v^1)$.

We can combine the above two mechanism , let the payment to the highest agent 1 is $p^1 = F(v^1, v^2)$, Such as $p^1 = (v^1 + v^2)/2$ . Obviously, $v^1 > p^1 > v^2$. In this mechanism, the higher $v^1$ is, the higher the $p^1$ is. As a result, it is less truthful than the Vichery Auction. In fact, it is partly truthful. The merit of this mechanism is that the utility of mechanism is bigger than that of Vichery Auction.

Further, we can let the $p^1 = F(v^1, v^2, \ldots, v^n)$.